\documentstyle{article}
\begin{document}
\begin{center}
{\Large\bf Causal Dissipative Cosmology  With Variable $G$ and $\Lambda$}
\end{center}
\vspace{2cm}
\begin{center}
{\bf Arbab I.Arbab}\footnote{ICTP, P.O. Box 586, 34100-Trieste, ITALY. E-mail:
arbab@ictp.trieste.it/ arbab64@hotmail.com} and {\bf A.Beesham}\footnote{Department
of Applied Mathematics, University of Zululand, Private Bag X1001, Kwa-Dlangezwa 3886, South Africa}
\end{center}
\vspace{4cm}
\begin{abstract}
\large
A cosmological model with variable $G$ and $\Lambda$ is considered in the framework of
Israel-Stewart-Hiscock(ISH) causal theory. 
Power law as well as inflationary solutions are
obtained.The gravitational constant is found to increase with time.
\end{abstract}
\vspace{2cm}
\newpage
\large
The role of dissipative effects on the evolution of the universe has been 
investigated by many works [6,8,11,15].
Weinberg has stressed that the entropy of the early universe may have emerged from bulk 
viscosity dissipation [12]. The bulk viscosity may arise from the conversion of massive 
string modes to massless modes as well as particle creation effects around the grand
unification (GUT) era [6].
It has been shown by Murphy and Barrow[8,11] that bulk viscosity can lead to 
inflation. In a recent paper [10], we have shown that this solution also arises 
in the presence of variable gravitational and cosmological constant. In the same
view Singh {\it et al} [13] obtained a similar solution with different energy
conservation.
Most of these investigation are based on Eckart theory. However, it is shown that
this theory is non-causal and all of it equilibrium states are unstable. 
The presently available theory considering dissipative 
effects in the universe is the full Israel-Stewart-Hiscock casual thermodynamics
[2,3]. 
In this paper we would like to investigate the evolution of bulk viscous 
stress in this casual theory where both $G$ and $\Lambda$ vary with time. 
Our investigation shows that the variation of the gravitational constant is also
consistent with the full casual theory. For age values consistent with present 
data the theory allows only increasing $G$ for the present phase. 
We further show that the functional dependence of
the coefficient of bulk viscosity assumed by many workers is an additional 
ad hoc assumption. We have found power law as well as inflationary solutions in the
full casual theory.
\\
{\bf The Field Equations}
\\
In a Robertson Walker(RW) flat metric, the Einstein field equations with a perfect fluid 
energy-momentum tenser and variable $G$ and $\Lambda$ yield
\begin{equation}
3\frac{\dot R^2}{R^2}=8\pi G\rho+\Lambda\  \ ,
\end{equation}
\begin{equation}
3\frac{\ddot R}{R}=-4\pi G(3p+\rho)+\Lambda
\end{equation}
and
\begin{equation}
3(p+\rho)\dot R=-(\frac{\dot G}{G}\rho+\dot\rho+\frac{\dot\Lambda}{8\pi G})R\ .
\end{equation}

The usual energy-momentum conservation, $T^{\mu\nu}_{;\nu}=0$, leads to
\begin{equation}
\dot\rho+3\frac{\dot R}{R}(p+\rho)=0\ .
\end{equation}
In the presence of bulk viscous stress ($\Pi$), the pressure term becomes
$p\rightarrow p+\Pi$.
\\
{\bf Section A}
\\
Equation (3) and (4) yields
  \begin{equation}
  \dot\rho+3H(\rho+p)=-3\Pi H
  \end{equation}
  and
  \begin{equation}
  8\pi \dot G\rho+\dot\Lambda=0\ .
  \end{equation}
  where a dot denotes a derivative with respect to time $t$.
   The density  $\rho$ and the pressure $p$ are related by the
   equation of the state
 \begin{equation}
 p=(\gamma-1)\rho\ ,
 \end{equation}
where $\gamma$=constant. 
Following [10,13] we take
\begin{equation}
\Lambda=3\beta H^2\ .
\end{equation}
Using eq.(6), eqs.(1) \& (8) yield
\begin{equation}
G=BH^{-2\beta/(1-\beta)}, \ B=\rm const.\ ,
\end{equation}
\begin{equation}
\rho=DH^{2/(1-\beta)}, D=\rm const.
\end{equation}
while eq.(2) gives
\begin{equation}
\dot H+\frac{3\gamma(1-\beta)}{2}H^2=-4\pi G\Pi, \ \beta\ne1\ .
\end{equation}
The causal evolution equation for the bulk viscous pressure is given by [2,3]
\begin{equation}
\Pi+\tau\dot\Pi=-3\zeta H-\frac{\epsilon}{2}\tau\Pi(3H+\frac{\dot\tau}{\tau}-\frac{\dot\zeta}{\zeta}-
\frac{\dot T}{T}), 
\end{equation}
where $T$ is the temperature, $\zeta$ the bulk viscosity coefficient and $\tau$ the relaxation time.
Setting $\epsilon=0, 1$ gives the truncated and the full-ISH causal
theory, respectively.
Now consider the parameterization
\begin{equation}
\zeta=\tau\rho, \ \ \rm and\ \ T=T_0\rho^r\ ,
\end{equation}
where $r=\frac{\gamma-1}{\gamma}$, $T_0$ are constants.
Upon using eq.(9), eq.(11) yields
\begin{equation}
\dot H+\frac{3\gamma(1-\beta)}{2}H^2=-4\pi B H^{-2\beta/(1-\beta)}\Pi, \ \beta\ne1 \
.
\end{equation}
We will consider here two cases:
\\
{\bf Case (1):}\\
(a) $H=H_0=\rm const.$ and $\beta\ne1$, so that $R\propto \exp(H_0t)$, i.e., an
inflationary
solution.
It follows from eqs.(8)-(12) that
\begin{equation}
\Lambda=\rm cons., G=\rm const., \rho=\rm const., \Pi=\rm const.,
\end{equation}
and
\begin{equation}
\zeta=\rm const., \ \rm and\ \ \tau=\rm const.
\end{equation}
We see that the bulk viscosity remains constant during the inflationary phase. This
solution is obtained by Wolf for the non-causal theory and by Banerjee and Beesham
for the casual theory.\\
(b) $H=\rm const.$ and $\beta=1$: We see that
$\Lambda=3H^2, R\propto\exp(\sqrt{\frac{\Lambda}{3}}t), G=\rm const. \ \rm and
\ \ \rho=0$. From eqs.(5) and (13) it follows that $\Pi=0$, $\zeta=0$ and $T=0$,so 
we
get the perfect fluid solution. Therefore, during this de Sitter solution the bulk
viscosity vanishes. Also, if $\alpha=\frac{3\gamma(1-\beta)}{2}, \Pi=0$ and
we recover the model in [9]. \\
{\bf Case(2)}:\\
A power law form of $R$, i.e., $R=At^\alpha, A, \alpha=\rm consts.$. Applying this
in eq.(15) one obtains,
\begin{equation}
\Pi=-Nt^{-2/(1-\beta)}, \beta\ne1\ ,
\end{equation}
 where $N=\frac{[-2+3\gamma\alpha(1-\beta)]\alpha^{(1+\beta)/(1-\beta)}}{8\pi A}$.
Equations
(9) and (10) give
\begin{equation}
\rho=B't^{-2/(1-\beta)}, G=Ct^{2\beta/(1-\beta)}, C, B' \ \rm consts., \beta\ne1\ .
\end{equation}
From the above equation we see that $\Pi=-a\rho, a=\rm const.$ Using eqs.(12) and (17),
eq.(13) gives
\begin{equation}
\tau=Bt, B=\rm const.
\end{equation}
and
\begin{equation}
\zeta=\zeta_0t^{-(1+\beta)/(1-\beta)},\ \zeta_0=\rm const,\ \beta\ne1\ .
\end{equation}
It is clear that the viscosity is decreasing with time. This solution is obtained
by Desikan [5] for Brans-Dicke theory with constant deceleration parameter under
the framework of ISH casual theory. This is manifested in the replacement of [$\alpha/(1+b)]$ by
$-2\beta/(1-\beta)$. We see that these relations are valid irrespective of the
functional dependence form of $\zeta$ on $\rho$. We would like to mention that
Banerjee and Beesham followed a lengthy way to arrive at this result.
They, however, claimed that this solution is possible only for particular behavior
i.e., $\zeta\propto \rho^{1/2}$. In fact this solution is valid for all values of
[$q$, see eqs.(2.8) and (2.14)]
\\
{\bf Section B}\\

Another way of looking at eq.(3) is by assuming that
\begin{equation}
\dot\rho+3\gamma H\rho=0
\end{equation}
and
\begin{equation}
\frac{\dot G}{G}\rho+\frac{\dot\Lambda}{8\pi G}=-3\Pi H.
\end{equation}
Using eqs.(1) and (8) this can be written as
\begin{equation}(\frac{2}{1-\beta}\frac{\dot H}{H}+3\gamma H)\rho=-3\Pi H.
\end{equation}
We now assume a power law for $R$, i.e., $R=At^\alpha, A, \alpha\ \ \ \rm consts.$.
Equation (21) and (23) now yield
\begin{equation}
\Pi=-N\rho, \ \ \rm and \ \ \rho=Dt^{-3\alpha\gamma}, D=\rm const., \beta\ne 1 ,
\end{equation}
where $N=\frac{-2+3\alpha\gamma(1-\beta)}{3\alpha(1-\beta)}$. Using eqs.(13) and
(24), eq.(12) gives
\begin{equation}
\tau=Bt
\end{equation}
where $B^{-1}=3\alpha[\gamma(1-\frac{\epsilon}{2}(1+r))+(\frac{1}{N}-\frac{\epsilon}{2})].$\\
In section A, we obtained a similar relation for the relaxation time and upon using
eq.(13) we obtain
\begin{equation}
\zeta=\tau\rho=\eta_0t^{1-3\alpha\gamma}, \eta_0=\rm const.
\end{equation}
From eqs.(1), (8) and (24), the gravitational parameter is given by
\begin{equation}
G\propto t^{-2+3\alpha\gamma}
\end{equation}
Therefore, $G$ increases or decreases with time according to whether
$3\alpha\gamma>2 \ \rm or \ \ <2$.
However, $G$ remains constant in the pure radiation and matter epochs. We therefore,
see that the casual bulk effects perturb the usual FRW solution for the radiation
and matter phases.
For the present era, the gravitational parameter is increasing with time( to get
a reasonable age parameter for the universe). 
\\
{\bf  Inflationary Solution}
\\
This is obtained when $H=H_0=\rm const$, i.e., $R=R_0\exp(H_0t), R_0=\rm const$
In this case eq.(20) gives
\begin{equation}
\rho=D\exp(-3\gamma H_0t), D=\rm const
\end{equation}
so that
\begin{equation}
G=F\exp(3\gamma H_0t), F=\rm const.
\end{equation}
and from eq.(23)
\begin{equation}
\Pi=-\gamma\rho .
\end{equation}
 Applying this to eq.(12) we get
 \begin{equation}
\tau=\rm const \ \ \ and \ \ \ \zeta\propto\exp(-3\gamma H_0t)\ .
\end{equation}
Note that during inflation both $\zeta$ and $\rho$ decreases with time
exponentially.
We also note that this solution is applicable irrespective of the value of
$\epsilon$,
i.e., for $\epsilon=1,\ \rm or\ \  0$.
 \\
 {\bf Acknowledgement}
 \\
 AIA wishes to thank the FRD (South Africa) for support and the University of Zululand for hospitality.
 \\
 \newpage
{\bf  References}
 \\
1. N. Banerjee and A.Beesham, Pramana 46(1996)213\\
2. W.A.Hiscock and L.Lindblom, Ann. Phys.(NY) 100, (1976)310\\
3. W.Israel, J.Stewart, Physics Letter A58, (1976)213\\
4. P.C. Paul, S.Mukhrejee and A.Beesham,(University of Zululand Preprint-1997)\\
5. K.Desikan, Pramana, 45(1995)511\\
6. C.Wolf, S.-Afr. Tydskr, 14(1991)68\\
7. R.Maartens, Class. Quantum Gravit. 12(1995)1455\\
8. D.J. Barrow, Phy. Lett.B180(1986)335\\
9. A.I.Arbab, (University of Zululand Preprint-1997)\\
10. A.I.Arbab, Gen. Rel. Gravit.29(1997)61\\
11. G.Murphy, Phys. Rev.D8 (1973)4321\\
12. S.Weinberg, Gravitation and Cosmology (J.Wiely, NY, 1972)\\
13. T.Singh, A.Beesham and W.S.Mbokazi, Gen. Relat. Gravit.30(1998)573\\
14. C.Ekart, Phys. Rev.58(1940)919\\
15. O.Gron, Astr. Phys. Space Science.173(1990)191\\
\end{document}